\documentclass{aastex63}
\usepackage{boites}

\usepackage{upgreek,hyphenat,aas_macros,longtable,array}

\usepackage{graphicx}
\usepackage{bm}
\usepackage{hyperref}
\received{14 June 2020}
\revised{29 July 2020}
\accepted{4 August 2020}
\submitjournal{and accepted by the International Journal of Astrobiology}

\shorttitle{}
\shortauthors{Wright}

\begin{document}

\title{Planck Frequencies as Schelling Points in SETI}

\correspondingauthor{Jason Wright}
\email{astrowright@gmail.com}

\author[0000-0001-6160-5888]{Jason T.\ Wright}
\affil{Department of Astronomy \& Astrophysics and \\
Center for Exoplanets and Habitable Worlds and\\ 
Penn State Extraterrestrial Intelligence Center\\ 525 Davey Laboratory \\
The Pennsylvania State University \\
University Park, PA, 16802, USA}

\begin{abstract}

In SETI, when searching for ``beacons''---transmissions intended for us and meant to get our attention---one must guess the appropriate frequency to search by considering what frequencies would be universally obvious to other species. This is a well known concept in game theory, where such solutions to a non-communicative cooperative game (such as a mutual search) are called ``Schelling points.''

It is noteworthy, therefore, that when developing his eponymous units, Planck called them ``natural'' because they ``remain meaningful for all times and also for extraterrestrial and non-human cultures.''

Here, I apply Planck's suggestion in the context of Schelling points in SETI with a ``Planck Frequency Comb,'' constructed by multiplying the Planck energy by integer powers of the fine structure constant. This comb includes a small number of frequencies in regions of the electromagnetic spectrum where laser and radio SETI typically operates. Searches might proceed and individual teeth in the comb, or at many teeth at once, across the electromagnetic spectrum. Indeed, the latter strategy can be additionally justified by the transmitter's desire to signal at many frequencies at once, to improve the chances that the receiver will guess one of them correctly.

There are many arbitrary and anthropocentric choices in this comb's construction, and indeed one can construct several different frequency combs with only minor and arbitrary modifications. This suggests that it may be fruitful to search for signals arriving in frequency combs of arbitrary spacing. And even though the frequencies suggested here are only debatably ``better'' than others proposed, the addition of the Planck Frequency Comb to the list of ``magic frequencies'' can only help searches for extraterrestrial beacons.
\end{abstract}


\section{Introduction} \label{sec:intro}

\subsection{Schelling Points}
\label{sec:intro.Schelling}

SETI can be divided into two general search strategies: searches for ``beacons,'' and searches for signals or other signs of technology not intended to be found by us. The search for beacons can be more focused because signals intended to be found should be obvious, and so should be strong, easily distinguished from backgrounds or astrophysical sources, and occur at times, frequencies, and places that are simple to guess, even for another species \citep[see, e.g.][]{WrightAdHoc,WrightExoplanetsSETI}.

\citet{schelling1960strategy} established the idea of ``focal points'' (often called ``Schelling points'' today) as optimal strategies in a game in which  players must coordinate their activities to reach a common goal {\it without communicating}. His example involved players attempting to find each other in New York City, with the focal point locations being famous landmarks and the focal point time being noon. Schelling's insight was that while the game appears to be hopeless at first, there clearly exist {\it sub}optimal strategies (searching random places at random times), and that guessing at the other team's strategy would likely result in success.  He was correct: this game was actually played as part of an American television program in 2006 and the players successfully found each other within hours.\footnote{``Mission Impossible: In Search of Strangers in New York City,'' ABC Primetime, March 16, 2006.}

The application to beacons in SETI is clear; indeed \citet{schelling1960strategy} cited the work of \citet{cocconi_searching_1959} and their choice of the 21-cm line of hydrogen as an example of an interstellar focal point. Since then, many ``magic frequencies'' \citep{Tarter1980} and other Schelling points have been suggested in SETI, spanning a range of potential places, frequencies, signal forms, and times to search signals \citep[see][for some examples in frequency space]{Drake73,Kardashev79,Blair92,Blair93,Gindilis93,Weber95,MorrisonPhD,Narusawa18}.\footnote{After publication, I became aware of \citet{Sivaram2019}, who anticipated some of the ideas in this paper, in particular, using powers of the fine structure constant to reduce frequencies composed of fundamental constants to values more easily searched. Indeed, one of their frequencies corresponds to twice the frequency of the $n=5$ tooth of the Rydberg Frequency Comb below.}

In the early days of radio SETI, radio telescopes were limited in the bandwidth they could search for strong signals, so there was value in guessing which frequencies would be most likely to have signals. Today, broadband instrumentation at radio and other telescopes often allows one to search a wide range of frequencies, and indeed the Breakthrough Listen backend at the Green Bank Telescope \citep{MacMahon18} searches billions of frequencies at once. There is still value in reasoning which band is best to search in, however, since we cannot search the entire EM spectrum at once and must choose which kinds of telescopes, receivers, and data analysis to do in a search. One also has higher sensitivity to signals of a {\it particular} frequency than to signals of {\it any} frequency because of the ``look elsewhere'' effect. Blind searches in frequency space can thus be complemented by focused searches at particular frequencies, even within the same data set.

An essential component of identifying Schelling points is finding points of commonality. For instance, players in the game in New York City mentioned above found each other because they were all aware of the major New York City landmarks and the significance of noon as a special time of day.\footnote{They found each other on the observation deck of the Empire State Building (perhaps the world's most famous skyscraper, made famous as a place to meet in the films ``An Affair to Remember'' and ``Sleepless in Seattle'') and in Times Square (a popular tourist attraction and the site of the annual ``ball drop'' televised nationwide every New Year's Eve.) Indeed, these landmarks are so well known that the preceding parentheticals are largely unnecessary except to illustrate the importance of a shared cultural heritage to the players' success.} This complicates its application to SETI: we have very little we can be sure we have in common with an alien species.

\subsection{Planck's Natural Units}
\label{sec:intro.Planck}

\citet{PlanckUnits} established his now-famous set of natural units based on the fundamental physical constants $G$, $c$, and $\hbar$. In that work, \citeauthor{PlanckUnits} wrote:  ``It is interesting to note that with the help of the [above constants] it is possible to introduce units\ldots which\ldots remain meaningful for all times and also for extraterrestrial and non-human cultures, and therefore can be understood as 'natural units'"\footnote{Translation by Sabine Hossenfelder.} and that ``\ldots these units keep their values as long as the laws of gravitation, the speed of light in vacuum, and the two laws of thermodynamics hold; therefore they must, when measured by other intelligences with different methods, always yield the same.''\footnote{Translation by Michael Hippke.} Planck's identification of these units as points of commonality with extraterrestrial species thus leads to their use as Schelling points.

Here, I shall explore the application of Planck units to the concept of Schelling points in frequency space for electromagnetic SETI---specifically, the use of the inverse of the Planck time $t_P = \sqrt{\hbar G/c^5}$ as a presumed universal standard for transmission frequency.

\section{The Planck Frequency Comb}

\label{sec:comb}
\subsection[]{Powers of $\alpha$}
\label{sec:comb.multiples}
The most salient problem with using the Planck time as a frequency standard is that it is so short.  A photon with the Planck energy $E_P = \sqrt{\hbar c^5/G}$ has of order $10^{23}$ eV, which is far above the energy of the most energetic photons known. Indeed, if such photons can exist at all, they can interact with cosmic microwave background photons and produce particle-antiparticle pairs, meaning that space is actually opaque to them.

There is, however another physical constant fundamental to electromagnetic (and weak force) radiation: the charge of the electron $e$. Note that this constant is not particular to the electron: {\it all} charges in nature are integer multiples of $e$, with the exception of the quarks, and even those come in multiples of $e/3$. Combined with the constants used in Planck's units, one can construct the dimensionless fine structure constant $\alpha\approx 1/137$, expressed in electrostatic units as $\alpha = e^2/(c\hbar)$.  

One can therefore construct an array of logarithmically spaced frequencies---a ``frequency comb''\footnote{Typically, frequency combs in physics are evenly spaced in frequency; this comb is evenly spaced in log frequency. There is also a nice metaphor of a comb as an object used to sift through hair or sand to find something, reminiscent of the way SETI seeks to find needles in the ``Cosmic Haystack,'' \citep{Haystack} but that is not my meaning here.}---defined by the Planck energy multiplied by integer powers of the fine structure constant. 

This array will include photons at a large number of frequencies, some of them at convenient places across the electromagnetic spectrum. That is, one can write that there exists an array of natural photon energies 
\begin{equation}
    E = E_P \alpha^n
\end{equation}

\noindent corresponding to frequencies
\begin{equation}
    \nu= \frac{\alpha^n}{2\pi t_P}
\end{equation}
\noindent for many integer values of $n$. Equivalently, one can think of $\alpha$ as the ``base'' of a multiplicative counting system in Planck units. We can then define a dimensionless quantity $n$ describing a photon with energy $E$ or frequency $\nu$ or wavelength $\lambda$ as
\begin{equation}
    n = \log_\alpha{(E/E_P)} = \log_\alpha{(2\pi\nu t_P)} = \log_\alpha{(2\pi l_P/\lambda)}
\end{equation}
\noindent where $l_P = \sqrt{\hbar G / c^3}$ is the Planck length. We can then identify the relevant Schelling points as frequencies or energies for which $n$ is an integer. I list those frequencies with integer values of $n$ in Table~\ref{tab:Comb}.

\begin{deluxetable}{rrrl}
\tablecolumns{4}
\tablecaption{The Planck Frequency Comb}
\tablehead{\colhead{$n$} &\colhead{Wavelength} & \colhead{Energy/frequency} & \colhead{Notes}}
\startdata
0 &  & $1.22089 \times 10^{28}$ $\mathrm{eV}$  &  Planck energy \\
1 &  & $8.90927 \times 10^{25}$ $\mathrm{eV}$  &   \\
2 &  & $6.50141 \times 10^{23}$ $\mathrm{eV}$  &   \\
3 &  & $4.74431 \times 10^{21}$ $\mathrm{eV}$  &   \\
4 &  & $3.46209 \times 10^{19}$ $\mathrm{eV}$  &  Highest energy cosmic rays \\
5 &  & 252.641 $\mathrm{PeV}$  &   \\
6 &  & 1.84361 $\mathrm{PeV}$  &   \\
7 &  & 13.4535 $\mathrm{TeV}$  &   \\
8 &  & 98.1746 $\mathrm{GeV}$  &  Observable by {\it Fermi} \\
9 &  & 716.415 $\mathrm{MeV}$  &  Observable by {\it Fermi} \\
10 &  & 5.22793 $\mathrm{MeV}$  &  Observable by {\it Fermi} \\
11 &  & 38.1501 $\mathrm{keV}$  &  Observable by {\it Fermi} \\
12 & 44.5354 $\mathrm{\mathring{A}}$ & 278.395 $\mathrm{eV}$  &  Observable by {\it Chandra} \\
13 & 6102.96 $\mathrm{\mathring{A}}$ & 2.03154 $\mathrm{eV}$  &  Optical \\
14 & 83.6325 $\mathrm{\upmu}$ & 3.58464 $\mathrm{THz}$  &   \\
15 & 1.14607 $\mathrm{cm}$ & 26.1584 $\mathrm{GHz}$  &  Microwave K band \\
16 & 1.57052 $\mathrm{m}$ & 190.887 $\mathrm{MHz}$  &  Observable by EoR experiments \\
17 & 215.218 $\mathrm{m}$ & 1.39297 $\mathrm{MHz}$  &  Ionospheric cutoff \\
18 &  & 10.1650 $\mathrm{kHz}$  &   \\
19 &  & 74.1775 $\mathrm{Hz}$  &  In the frequency range of LIGO \\
20 &  & 541.300 $\mathrm{mHz}$  &   \\
21 &  & 3.95005 $\mathrm{mHz}$  &  In the frequency range of LISA \\
22 &  & 28.8249 $\mathrm{\upmu Hz}$  &   \\
23 &  & 210.346 $\mathrm{nHz}$  &  In the frequency range of PTAs \\
24 &  & 1.53497 $\mathrm{nHz}$  &  In the frequency range of PTAs \\

\enddata
\tablecomments{$n = \log_\alpha{(E/E_P)}$. \\
These values are uncertain at the level of 7 ppm, limited by our knowledge of $G$. This is comparable to the precision with which we know the frame of the CMB and LSR.\label{tab:Comb}}
\end{deluxetable}

For completeness, I include a wide range of frequencies here. I note in this and subsequent tables the frequency ranges of some space observatories capable of detecting these photons, and I mark the approximate energy of the highest energy cosmic rays (but these are presumably atomic nuclei, not photons).

Except in the midinfrared where airborne or space missions are necessary, from the optical through the radio there exist a wide array of ground-based telescopes that could perform spectroscopy to search for these signals. I note in the tables some of the optical/near-infrared bands, the frequency range of ALMA, the various microwave bands, and the frequencies typical of telescopes designed to detect the Epoch of Reionization \citep{Fulranetto06}.

Below a few MHz radio waves cannot penetrate the ionosphere, however space missions such as the Netherlands-China Low-Frequency Explorer and SunRISE are planned to observe this part of the EM spectrum. Past this point I list in the notes the approximate frequencies corresponding to the gravitational wave detection experiments at LIGO, the planned space gravitational wave antenna LISA, and the pulsar timing arrays (PTAs).  

Interestingly, there is a comb tooth in the optical, $n=13$ at 6103\AA.  Then next tooth, at $n=14$, has $\lambda=$83.632$\upmu$, and has a high background contamination from the Earth's atmosphere but is observable by the FIFI-LS instrument on {\it SOFIA} and was in the bandpass of {\it Herschel}. In the microwave, the $n=15$ tooth is at 26.158 GHz, in K band and accessible to many radio telescopes. The $n=16$ tooth at 190.89 MHz is subject to significant terrestrial radio frequency interference, but is accessible by, for instance, the Murchison Widefield Array and the Five hundred meter Aperture Spherical Telescope. There are no comb teeth in the frequency range spanned by ALMA.

\subsection{Uncertainties in Arrival Frequency}
\label{sec:comb.uncertainties}
The precision which one can predict the arrival frequencies of signals transmitted in a Planck Frequency Comb is limited by our knowledge of the fundamental constants, and the frame of the transmitter. 

According to CODATA\footnote{\url{https://physics.nist.gov/cuu/Constants/index.html}}, our uncertainty in the values of the Planck units is around 7ppm and dominated by the uncertainty in $G$.  Expressed as Doppler shifts, this frequency uncertainty corresponds to around 2 km/s. Our uncertainty in $\alpha$ is four orders of magnitude smaller.  

Any transmission will be Doppler shifted by the velocities of the transmitter and receiver.  One can correct our measurements for the motion of the receiver in the Solar System barycentric frame \citep[e.g.][]{bary}, but accelerations of the transmitter will cause the frequency of its transmissions to drift \citep[e.g.][]{Sheikh19}.  Even a transmitter that is not appreciably accelerating or that is correcting for its acceleration will presumably have some nonzero relative velocity to the Solar System barycenter.

A transmitter operating at a predictable frequency would therefore need to adjust their transmission frequencies to that of some universal frame of reference. There are a few popular choices in the literature for such a universal frame (itself a Schelling point). 

The most literally universal frame is that of the cosmic microwave background (CMB). The precise velocity of this frame is somewhat uncertain because our measurement of it is largely degenerate with the $l=1$ CMB anisotropies, which are subject to cosmic variance. It is thus unclear whether the ``best'' Schelling point here is the {\it true} frame of the CMB, or the frame in which the dipole anisotropy is measured to be zero, which is much easier to determine precisely. As a practical matter, however, the difference between these two frames is smaller than our measurement error in the dipole, which corresponds to around 1 km/s  \citep{Planck14}.  

Alternatively, one might choose the Galactic barycenter, the Local Standard of Rest (LSR), or the solar system barycenter (SSB) as the relevant frame. The Galactic barycenter makes sense if the transmitter is also within the Milky Way, or if the signal is being broadcast to the entire Galaxy. The LSR is appropriate for signals originating in or targeting nearby stars, and the SSB would be appropriate for signals targeting the solar system, specifically. \citet{horowitz93}, for instance, checked the CMB, Galactic barycenter, and SSB frames.

Our uncertainty in the velocity of the LSR is of order 1 km/s \citep{Schonrich10}, while our uncertainty in the velocity of the sun in the frame of the Galactic barycenter is of order 5 km/s \citep{Reid14}. Our knowledge of the SSB frame is exquisite, at least six orders of magnitude smaller than this.

The arrival frequencies of the Planck Frequency Comb are thus uncertain by 2--5 km/s (7--18 ppm), depending on the frame in which one searches for them.

\section{Anthropecentrism of ``Natural'' Units, and Alternative Formulations}
\label{sec:alternatives}

Planck felt that his units were ``natural,'' meaning that they transcended not just the physics tradition he was trained in, but humanity itself because they used only fundamental constants. This makes sense: we refer to these constants of nature as ``fundamental'' because they describe properties of space, time, and the fundamental forces, as opposed to the properties of the {\it content} of the universe (atomic constants, for instance) which can, in principle at least, be derived from them. One can easily imagine that all physics traditions, including extraterrestrial ones, would come to similar conclusions.

However, one can imagine many other ways to formulate the construction of natural units and the fundamental constants, and this complicates their use as Schelling points.  To give a few examples:
\begin{itemize}
    \item Planck formulated his constants in terms of $\hbar$, so that the inverse of the Planck time corresponds to an {\it angular} frequency, which is a common convention in physics. But astronomers, for instance, typically express the energetic quality of light in cycles per second or wavelength, and so prefer to write $E=h\nu$ or $E=hc/\lambda$ instead of $E=\hbar\omega$. This feels awkward to some physicists because of the way we formulate physics in terms of algebra and our desire to be parsimonious in the number of symbols we use (avoiding lots of ``2$\pi$''s flying around) but the traditions of physics and engineering of other species may teach differently. One should therefore consider a set of natural units and $\alpha$ formulated with $h$ instead of $\hbar$, and this would yield a different set of frequencies.
    \item When determining the ``base'' by which to multiply the Planck units, instead of the fine structure constant (which uses $e$, a constant of nature not found in the Planck formulation) one could use the base of the natural logarithm (also $e$!). One could, of course, use {\it any} dimensionless number.
    \item In formulating the fine structure constant, one could also choose the smallest unit of charge in nature, that of the quark, $e/3$, instead of the charge of the electron.
    \item The dark energy appears to provide an additional fundamental scale to the universe.  Its energy density of $\sim 7\times 10^{-30}$g/cm$^3$ can be used to construct a frequency known to any species with an understanding of cosmology (roughly $10^{-19}$ Hz, comparable to the present-day value of the Hubble constant).  Our current best measurements of this quantity are probably not sufficiently precise for its use as a Schelling point in frequency, however.
    \item One can multiply any apparently natural constant by an arbitrary dimensionless constant and have an equally valid constant. It is possible and perhaps likely that another species would have additional factors of 2, 3 or even $\pi$ in their formulations of ``fundamental'' physical constants.
    \item For instance, even our use of $\pi$ as a fundamental constant of mathematics is somewhat arbitrary: much or most of mathematics and physics can be expressed in a smaller number of algebraic symbols with the substitution $2\pi\rightarrow\tau$.\footnote{As only half-jokingly illustrated by Michael Hartl's ``Tauist manifesto.'' \url{https://web.archive.org/web/20200419074326/https://tauday.com/tau-manifesto}}
    \item Other species may find it more obvious to transmit and receive at noninteger values of $n$, for instance half-integers or powers of $2\pi$.
    \item The use of the electron charge $e$ in constructing the Planck Frequency Comb is most self-consistent when used to measure the frequency of photons (or neutrinos), but in principle one could use these same frequencies as a Schelling point for a beacon in gravitational waves. Alternatively, one might in this case use instead the gravitational fine structure constant, $\alpha_G = (m_e/m_P)^2 \approx 1.752\times10^{-45}$ where $m_e$ is the electron mass and $m_P$ is the Planck mass. In additional to somewhat arbitrarily using the electron's mass in a Schelling point for gravitational waves, it produces comb spacings so wide that there is barely even a single useful frequency: $n=1$ corresponds to $\nu = 5.1715$ mHz.  
    \item There are other constants of nature one could use. The detection of electromagnetic radiation always involves interactions with electrons, so it might be obvious to use $m_e$ as a basis for a frequency comb. Indeed, the comb defined by $E=(\frac{1}{2}m_ec^2)\alpha^n$ contains another well-known unit of energy at $n=2$: the Rydberg ($R_\infty = $13.6 eV, almost exactly the ionization potential of ground-state hydrogen).
\end{itemize}

Making different choices than Planck and I have made amount to changing the constant factor ($2\pi/t_P$) and scaling factor ($\alpha$) of the Planck Frequency Comb. Because of this, the use of natural units as a Schelling point is perhaps best applied as a search for combs of signals of {\it arbitrary} constant and spacing, removing the need for one to guess at another species' preference for the above choices.  This complicates efforts to search a particular frequency, but implies that beacons may transmit at more than one frequency, corresponding to multiple teeth of their comb (so that the constant and spacing can be identified as obviously artificial and tied to the physical constants). 

I provide in Tables~\ref{tab:halpha}--\ref{tab:Ry} of the Appendix lists of Planck frequencies for different choices of $\hbar$ vs.\ $h$ (in both $E_P$ and $\alpha$, for consistency), for the natural logarithm vs.\ $\alpha$, and also the ``Rydberg Frequency Comb.''  To distinguish these alternative schemes from the one in table~\ref{tab:Comb}, one can index $n$ by which version of Planck's constant and which base it uses, yielding:
\begin{eqnarray}
 n \equiv n_{\hbar,\alpha} &=& \frac{\log{(E/E_P)}}{\log{\alpha}} \\
 n_{\hbar,e} &=& -\ln{(E/E_P)} \\
 n_{h,\alpha} &=& \frac{\log{(E/\sqrt{hc^5/G})}}{\log{(e^2/(ch))}} \\
 n_{h,e} &=& -\ln{(E/\sqrt{hc^5/G})} \\
 n_{R} &=& \frac{\log{(E/(m_ec^2/2))}}{\log{\alpha}} = \frac{\log{(E/R_\infty)}}{\log{\alpha}}-2
\end{eqnarray}

Significant wavelengths and frequencies appearing in these tables include 6332\AA\ and 6867\AA\ in the optical, 1.721$\upmu$ and 1.867$\upmu$ in H band, and 1.07 GHz, 1.278 GHz, 2.528 GHz, 2.683 GHz, 2.909 GHz, and 7.908 GHz in the microwave. 

Doubtless, other formulations of the Planck Frequency Comb can be made. Indeed, one might equally strongly argue that the empirically observed frequencies of strong atomic and molecular lines are more obvious as Schelling points than those constructed using abstract fundamental constants, and that a beacon would multiply these by ``important'' mathematical constants to tellingly distinguish them from astrophysical sources \citep[``$\pi$ times hydrogen,'' for instance. See][for a list of examples.]{Blair93} Put another way, to an observational astronomer the frequencies nature provides observationally are the most obvious bases for units because they imagine their alien counterparts using radio telescopes observe the same thing, while for a theoretical physicist it is the fundamental constants of nature that provide that commonality because to build such a transmitter requires an understanding of physics.

Despite these difficulties, the Planck Frequency Comb I present in Table~\ref{tab:Comb} is, I think, closely aligned with the spirit of  Planck's original suggestion and among the most parsimonious in terms of number of algebraic symbols and physical constants.

But the number of choices one needs to make to construct the comb and choose a reference frame---and the fact that other physics traditions might use different units or formulations of physics or choices of frame ---highlights the difficulty in applying Schelling's insight to SETI. One wishes to find certain points of commonality that are not specific to humans, but it can be challenging to identify which aspects of our physics traditions are truly universal, which are particular to how humans think, and which are simply accidents of how our physics has developed. 

It is possible that there is some way to measure the parsimony of expression of quantities that would be present in all physics and mathematical traditions, including extraterrestrial ones, and so help guide work identifying such Schelling points. Such a possibility could be worthwhile exploring as an interdisciplinary effort among mathematics, physics, complexity theory, and anthropology. 

\section{Applications}

 The most straightforward application of the Planck Frequency Comb as a Schelling point is to search for beacons at the frequencies of its teeth.  

The importance of identifying specific search frequencies is not as important as it once was, because modern astronomical spectroscopy can search a large number of frequencies simultaneously. For instance, the Breakthrough Listen backend \citep{MacMahon18} has a bandwidth of over 6 GHz, allowing it to perform high resolution spectroscopy across the entirety of any of the Green Bank Telescope's lower frequency receivers in a single integration.  

Setting aside the issues of anthropogenic radio interference and background from the Earth's atmosphere, the sensitivity of such work to a narrowband signal is in principle set by the noise of the instrument and the length of the integration \citep{siemion13}.  The threshold one chooses to identify candidate detections can then be set by the number of candidates one wishes to screen or, almost equivalently, the probability that signal of a given strength would have been observed by chance. In either case, the threshold is a function of the number of independent frequencies one observes. 

For instance, the Breakthrough Listen radio search \citep[e.g.][]{Enriquez17,Price20,SheikhETZ} searches billions of 2.7 Hz channels simultaneously, and typically sets a 25-$\sigma$ threshold for detection above the instrumental noise. The high-resolution optical continuous-wave laser search of  \citet{Tellis15} and \citet{Tellis17} had a bandwidth of a factor of 2 (from 3640--7890\AA) and a resolution of 60,000, meaning they searched $\sim$45,000 independent frequency channels simultaneously. They chose a threshold equivalent to $\sim 7$-$\sigma$ to generate a manageable number of candidates.

Application of a ``magic frequency'' search using the Planck Frequency Comb or other list of frequencies allows for a much more sensitive search, because the number of trial frequencies is considerably lower.  For instance, there are only 2 or 3 frequencies in the tables in this work in a given radio or optical band, meaning that they could each be examined individually for signals at the noise limit without any preliminary candidate thresholding, improving the sensitivity of the above searches by factors of several.

Another application would be to search for signals transmitted at multiple comb teeth, especially simultaneously. For the frequencies suggested in this work, this would usually involve observations with different kinds of telescopes or receivers, and then combining the significance of any signals found at multiple tooth frequencies. One might do such observations simultaneously, or one could prioritize observations by following up candidate signals at one comb tooth with observations at another. Indeed, there is little reason to suppose that a species would choose only a single mode of communication. Especially if success requires the receiver to guess the transmitter's frequency, transmitting all along an comb of frequencies makes strategic sense as a way to maximize the chances that they will guess correctly.

One could also acknowledge the uncertainty in correctly guessing the constant and spacing of the comb, and instead perform a search similar to that above but for all possible combs, including tightly-spaced combs with many teeth in the spectral grasp of single broadband instrument. This would be similar to the approach of \citet{Borra2012} who advocated searching for light from laser frequency combs (which would evenly spaced in frequency, not log frequency as in the Planck Frequency Comb). 

\section{Conclusions}

I have developed the suggestion of \citet{PlanckUnits} that his natural units would be recognizable to extraterrestials and the insight of \citet{schelling1960strategy} that such commonalities are useful for determining frequencies in SETI, to produce lists of frequencies expressible entirely in terms of fundamental constants and small integers.  Specifically, by multiplying the Planck energy by powers of the fine structure constant, I have constructed a ``Planck Frequency Comb'' of these special frequencies and suggest that they receive extra attention in SETI. 

Significant teeth in the comb include 6103\AA, 83.632$\upmu$, 26.158 GHz, and 190.89 MHz. There is also a single such frequency using the gravitational fine structure constant instead of the electromagnetic one: 5.1715 mHz, which is within the frequency range of LISA. The Planck Frequency Comb thus provides a set of frequencies to search across the electromagnetic spectrum (and beyond).

This analysis also suggests two additional search modalities: searching for signals in multiple channels simultaneously, for instance in the optical and the radio, at frequencies corresponding to multiple teeth of the same comb; and searching for combs of arbitrary spacing, for instance within the spectral grasp of a single instrument.

I have acknowledged and explored the somewhat arbitrary choices in the construction of these frequencies that reflect the idiosyncrasies of how we formulate physics, and provide four alternative sets of frequencies that use the unreduced Planck constant $h$ instead of $\hbar$, the base of the natural logarithm instead of $\alpha$, or the Rydberg instead of the Planck energy. An objective and, hopefully, universal model of mathematical parsimony might help guide future work in identifying appropriate ``magic frequencies.'' 

In the meantime, there does not seem to be a very strong argument that the Planck Frequency Comb provides a superior set of Schelling points to the ``magic frequencies'' already proposed in the literature, and indeed there are likely a very large number of similarly compelling frequencies that have not been proposed yet. This does not mean, however, that proposing new magic frequencies is a fruitless exercise. Much like how the teams looking for each other in New York City were better served visiting various Manhattan landmarks than searching randomly, it can only help to add the frequencies of the Planck Frequency Comb to the list of proposed Schelling points.

\acknowledgements{I thank Jos\'e Alberto Rubi\~no for pointing me to the Planck result for the true frame of the CMB, Michael Hippke for discussions about applications to gravitational waves,  Andrew Siemion for references for ``magic frequencies,'' and Claire Webb for many helpful suggestions that improved this paper. I thank Sabine Hossenfelder and Michael Hippke for their translations of \citet{PlanckUnits}. I thank the referee for helpful suggestions.

The Center for Exoplanets and Habitable Worlds and the Penn State Extraterrestrial Intelligence Center are supported by the Pennsylvania State University and the Eberly College of Science.

This research has made use of NASA's Astrophysics Data System Bibliographic Services and Astropy,\footnote{\url{http://www.astropy.org}} a community-developed core Python package for Astronomy \citep{2013A&A...558A..33A,2018AJ....156..123A}.}



\begin{thebibliography}{}

\bibitem[{Astropy Collaboration} et~al., 2018]{2018AJ....156..123A}
{Astropy Collaboration}, {Price-Whelan}, A.~M., {Sip{\H o}cz}, B.~M.,
  {G{\"u}nther}, H.~M., {Lim}, P.~L., {Crawford}, S.~M., {Conseil}, S.,
  {Shupe}, D.~L., {Craig}, M.~W., {Dencheva}, N., {Ginsburg}, A., {VanderPlas},
  J.~T., {Bradley}, L.~D., {P{\'e}rez-Su{\'a}rez}, D., {de Val-Borro}, M.,
  {Aldcroft}, T.~L., {Cruz}, K.~L., {Robitaille}, T.~P., {Tollerud}, E.~J.,
  {Ardelean}, C., {Babej}, T., {Bach}, Y.~P., {Bachetti}, M., {Bakanov}, A.~V.,
  {Bamford}, S.~P., {Barentsen}, G., {Barmby}, P., {Baumbach}, A., {Berry},
  K.~L., {Biscani}, F., {Boquien}, M., {Bostroem}, K.~A., {Bouma}, L.~G.,
  {Brammer}, G.~B., {Bray}, E.~M., {Breytenbach}, H., {Buddelmeijer}, H.,
  {Burke}, D.~J., {Calderone}, G., {Cano Rodr{\'{\i}}guez}, J.~L., {Cara}, M.,
  {Cardoso}, J.~V.~M., {Cheedella}, S., {Copin}, Y., {Corrales}, L.,
  {Crichton}, D., {D'Avella}, D., {Deil}, C., {Depagne}, {\'E}., {Dietrich},
  J.~P., {Donath}, A., {Droettboom}, M., {Earl}, N., {Erben}, T., {Fabbro}, S.,
  {Ferreira}, L.~A., {Finethy}, T., {Fox}, R.~T., {Garrison}, L.~H., {Gibbons},
  S.~L.~J., {Goldstein}, D.~A., {Gommers}, R., {Greco}, J.~P., {Greenfield},
  P., {Groener}, A.~M., {Grollier}, F., {Hagen}, A., {Hirst}, P., {Homeier},
  D., {Horton}, A.~J., {Hosseinzadeh}, G., {Hu}, L., {Hunkeler}, J.~S.,
  {Ivezi{\'c}}, {\v Z}., {Jain}, A., {Jenness}, T., {Kanarek}, G., {Kendrew},
  S., {Kern}, N.~S., {Kerzendorf}, W.~E., {Khvalko}, A., {King}, J., {Kirkby},
  D., {Kulkarni}, A.~M., {Kumar}, A., {Lee}, A., {Lenz}, D., {Littlefair},
  S.~P., {Ma}, Z., {Macleod}, D.~M., {Mastropietro}, M., {McCully}, C.,
  {Montagnac}, S., {Morris}, B.~M., {Mueller}, M., {Mumford}, S.~J., {Muna},
  D., {Murphy}, N.~A., {Nelson}, S., {Nguyen}, G.~H., {Ninan}, J.~P.,
  {N{\"o}the}, M., {Ogaz}, S., {Oh}, S., {Parejko}, J.~K., {Parley}, N.,
  {Pascual}, S., {Patil}, R., {Patil}, A.~A., {Plunkett}, A.~L., {Prochaska},
  J.~X., {Rastogi}, T., {Reddy Janga}, V., {Sabater}, J., {Sakurikar}, P.,
  {Seifert}, M., {Sherbert}, L.~E., {Sherwood-Taylor}, H., {Shih}, A.~Y.,
  {Sick}, J., {Silbiger}, M.~T., {Singanamalla}, S., {Singer}, L.~P., {Sladen},
  P.~H., {Sooley}, K.~A., {Sornarajah}, S., {Streicher}, O., {Teuben}, P.,
  {Thomas}, S.~W., {Tremblay}, G.~R., {Turner}, J.~E.~H., {Terr{\'o}n}, V.,
  {van Kerkwijk}, M.~H., {de la Vega}, A., {Watkins}, L.~L., {Weaver}, B.~A.,
  {Whitmore}, J.~B., {Woillez}, J., {Zabalza}, V., and {Astropy Contributors}
  (2018).
\newblock {The Astropy Project: Building an Open-science Project and Status of
  the v2.0 Core Package}.
\newblock {\em \aj}, 156:123.

\bibitem[{Astropy Collaboration} et~al., 2013]{2013A&A...558A..33A}
{Astropy Collaboration}, {Robitaille}, T.~P., {Tollerud}, E.~J., {Greenfield},
  P., {Droettboom}, M., {Bray}, E., {Aldcroft}, T., {Davis}, M., {Ginsburg},
  A., {Price-Whelan}, A.~M., {Kerzendorf}, W.~E., {Conley}, A., {Crighton}, N.,
  {Barbary}, K., {Muna}, D., {Ferguson}, H., {Grollier}, F., {Parikh}, M.~M.,
  {Nair}, P.~H., {Unther}, H.~M., {Deil}, C., {Woillez}, J., {Conseil}, S.,
  {Kramer}, R., {Turner}, J.~E.~H., {Singer}, L., {Fox}, R., {Weaver}, B.~A.,
  {Zabalza}, V., {Edwards}, Z.~I., {Azalee Bostroem}, K., {Burke}, D.~J.,
  {Casey}, A.~R., {Crawford}, S.~M., {Dencheva}, N., {Ely}, J., {Jenness}, T.,
  {Labrie}, K., {Lim}, P.~L., {Pierfederici}, F., {Pontzen}, A., {Ptak}, A.,
  {Refsdal}, B., {Servillat}, M., and {Streicher}, O. (2013).
\newblock {Astropy: A community Python package for astronomy}.
\newblock {\em \aap}, 558:A33.

\bibitem[{Blair} et~al., 1992]{Blair92}
{Blair}, D.~G., {Norris}, R.~P., {Troup}, E.~R., {Twardy}, R., {Wellington},
  K.~J., {Williams}, A.~J., {Wright}, A.~E., and {Zadnik}, M.~G. (1992).
\newblock {A narrow-band search for extraterrestrial intelligence (SETI) using
  the interstellar contact channel hypothesis}.
\newblock {\em \mnras}, 257:105--109.

\bibitem[Blair and Zadnik, 1993]{Blair93}
Blair, D.~G. and Zadnik, M.~G. (1993).
\newblock {A list of possible interstellar communication channel frequencies
  for SETI}.
\newblock {\em Astronomy and Astrophysics}, 278:669--672.

\bibitem[{Borra}, 2012]{Borra2012}
{Borra}, E.~F. (2012).
\newblock {Searching for Extraterrestrial Intelligence Signals in Astronomical
  Spectra, Including Existing Data}.
\newblock {\em \aj}, 144(6):181.

\bibitem[Cocconi and Morrison, 1959]{cocconi_searching_1959}
Cocconi, G. and Morrison, P. (1959).
\newblock Searching for interstellar communications.
\newblock {\em Nature}, 184(4690):844--846.

\bibitem[{Drake} and {Sagan}, 1973]{Drake73}
{Drake}, F.~D. and {Sagan}, C. (1973).
\newblock {Interstellar Radio Communication and the Frequency Selection
  Problem}.
\newblock {\em \nat}, 245(5423):257--258.

\bibitem[{Enriquez} et~al., 2017]{Enriquez17}
{Enriquez}, J.~E., {Siemion}, A., {Foster}, G., {Gajjar}, V., {Hellbourg}, G.,
  {Hickish}, J., {Isaacson}, H., {Price}, D.~C., {Croft}, S., {DeBoer}, D.,
  {Lebofsky}, M., {MacMahon}, D.~H.~E., and {Werthimer}, D. (2017).
\newblock {The Breakthrough Listen Search for Intelligent Life: 1.1-1.9 GHz
  Observations of 692 Nearby Stars}.
\newblock {\em \apj}, 849:104.

\bibitem[{Furlanetto} et~al., 2006]{Fulranetto06}
{Furlanetto}, S.~R., {Oh}, S.~P., and {Briggs}, F.~H. (2006).
\newblock {Cosmology at low frequencies: The 21 cm transition and the
  high-redshift Universe}.
\newblock {\em \physrep}, 433(4-6):181--301.

\bibitem[Gindilis et~al., 1993]{Gindilis93}
Gindilis, L., Davydov, V., and Strelnitski, V. (1993).
\newblock {New "Magic" Frequencies for SETI}.
\newblock In Shostak, G.~S., editor, {\em Third Decennial US-USSR Conference on
  SETI}, pages 161--163.

\bibitem[{Horowitz} and {Sagan}, 1993]{horowitz93}
{Horowitz}, P. and {Sagan}, C. (1993).
\newblock {Five years of Project META - an all-sky narrow-band radio search for
  extraterrestrial signals}.
\newblock {\em \apj}, 415:218--235.

\bibitem[{Kardashev}, 1979]{Kardashev79}
{Kardashev}, N.~S. (1979).
\newblock {Strategy for the search for extraterrestrial intelligence}.
\newblock {\em Acta Astronautica}, 6:33--46.

\bibitem[{MacMahon} et~al., 2018]{MacMahon18}
{MacMahon}, D.~H.~E., {Price}, D.~C., {Lebofsky}, M., {Siemion}, A.~P.~V.,
  {Croft}, S., {DeBoer}, D., {Enriquez}, J.~E., {Gajjar}, V., {Hellbourg}, G.,
  {Isaacson}, H., {Werthimer}, D., {Abdurashidova}, Z., {Bloss}, M., {Brandt},
  J., {Creager}, R., {Ford}, J., {Lynch}, R.~S., {Maddalena}, R.~J.,
  {McCullough}, R., {Ray}, J., {Whitehead}, M., and {Woody}, D. (2018).
\newblock {The Breakthrough Listen Search for Intelligent Life: A Wideband Data
  Recorder System for the Robert C. Byrd Green Bank Telescope}.
\newblock {\em \pasp}, 130(4):044502.

\bibitem[{Morrison}, 2017]{MorrisonPhD}
{Morrison}, I. (2017).
\newblock Constraining the discovery space for artificial interstellar signals.

\bibitem[{Narusawa} et~al., 2018]{Narusawa18}
{Narusawa}, S.-y., {Aota}, T., and {Kishimoto}, R. (2018).
\newblock {Which colors would extraterrestrial civilizations use to transmit
  signals?: The ``magic wavelengths'' for optical SETI}.
\newblock {\em \na}, 60:61--64.

\bibitem[{Planck}, 1900]{PlanckUnits}
{Planck}, M. (1900).
\newblock {Ueber irreversible Strahlungsvorg{\"a}nge}.
\newblock {\em Annalen der Physik}, 306(1):69--122.

\bibitem[{Planck Collaboration} et~al., 2014]{Planck14}
{Planck Collaboration}, {Aghanim}, N., {Armitage-Caplan}, C., {Arnaud}, M.,
  {Ashdown}, M., {Atrio-Barandela}, F., {Aumont}, J., {Baccigalupi}, C.,
  {Banday}, A.~J., {Barreiro}, R.~B., {Bartlett}, J.~G., {Benabed}, K.,
  {Benoit-L{\'e}vy}, A., {Bernard}, J.~P., {Bersanelli}, M., {Bielewicz}, P.,
  {Bobin}, J., {Bock}, J.~J., {Bond}, J.~R., {Borrill}, J., {Bouchet}, F.~R.,
  {Bridges}, M., {Burigana}, C., {Butler}, R.~C., {Cardoso}, J.~F., {Catalano},
  A., {Challinor}, A., {Chamballu}, A., {Chiang}, H.~C., {Chiang}, L.~Y.,
  {Christensen}, P.~R., {Clements}, D.~L., {Colombo}, L.~P.~L., {Couchot}, F.,
  {Crill}, B.~P., {Curto}, A., {Cuttaia}, F., {Danese}, L., {Davies}, R.~D.,
  {Davis}, R.~J., {de Bernardis}, P., {de Rosa}, A., {de Zotti}, G.,
  {Delabrouille}, J., {Diego}, J.~M., {Donzelli}, S., {Dor{\'e}}, O., {Dupac},
  X., {Efstathiou}, G., {En{\ss}lin}, T.~A., {Eriksen}, H.~K., {Finelli}, F.,
  {Forni}, O., {Frailis}, M., {Franceschi}, E., {Galeotta}, S., {Ganga}, K.,
  {Giard}, M., {Giardino}, G., {Gonz{\'a}lez-Nuevo}, J., {G{\'o}rski}, K.~M.,
  {Gratton}, S., {Gregorio}, A., {Gruppuso}, A., {Hansen}, F.~K., {Hanson}, D.,
  {Harrison}, D.~L., {Helou}, G., {Hildebrandt}, S.~R., {Hivon}, E., {Hobson},
  M., {Holmes}, W.~A., {Hovest}, W., {Huffenberger}, K.~M., {Jones}, W.~C.,
  {Juvela}, M., {Keih{\"a}nen}, E., {Keskitalo}, R., {Kisner}, T.~S., {Knoche},
  J., {Knox}, L., {Kunz}, M., {Kurki-Suonio}, H., {L{\"a}hteenm{\"a}ki}, A.,
  {Lamarre}, J.~M., {Lasenby}, A., {Laureijs}, R.~J., {Lawrence}, C.~R.,
  {Leonardi}, R., {Lewis}, A., {Liguori}, M., {Lilje}, P.~B.,
  {Linden-V{\o}rnle}, M., {L{\'o}pez-Caniego}, M., {Lubin}, P.~M.,
  {Mac{\'\i}as-P{\'e}rez}, J.~F., {Mandolesi}, N., {Maris}, M., {Marshall},
  D.~J., {Martin}, P.~G., {Mart{\'\i}nez-Gonz{\'a}lez}, E., {Masi}, S.,
  {Massardi}, M., {Matarrese}, S., {Mazzotta}, P., {Meinhold}, P.~R.,
  {Melchiorri}, A., {Mendes}, L., {Migliaccio}, M., {Mitra}, S., {Moneti}, A.,
  {Montier}, L., {Morgante}, G., {Mortlock}, D., {Moss}, A., {Munshi}, D.,
  {Naselsky}, P., {Nati}, F., {Natoli}, P., {N{\o}rgaard-Nielsen}, H.~U.,
  {Noviello}, F., {Novikov}, D., {Novikov}, I., {Osborne}, S., {Oxborrow},
  C.~A., {Pagano}, L., {Pajot}, F., {Paoletti}, D., {Pasian}, F., {Patanchon},
  G., {Perdereau}, O., {Perrotta}, F., {Piacentini}, F., {Pierpaoli}, E.,
  {Pietrobon}, D., {Plaszczynski}, S., {Pointecouteau}, E., {Polenta}, G.,
  {Ponthieu}, N., {Popa}, L., {Pratt}, G.~W., {Pr{\'e}zeau}, G., {Puget},
  J.~L., {Rachen}, J.~P., {Reach}, W.~T., {Reinecke}, M., {Ricciardi}, S.,
  {Riller}, T., {Ristorcelli}, I., {Rocha}, G., {Rosset}, C.,
  {Rubi{\~n}o-Mart{\'\i}n}, J.~A., {Rusholme}, B., {Santos}, D., {Savini}, G.,
  {Scott}, D., {Seiffert}, M.~D., {Shellard}, E.~P.~S., {Spencer}, L.~D.,
  {Sunyaev}, R., {Sureau}, F., {Suur-Uski}, A.~S., {Sygnet}, J.~F., {Tauber},
  J.~A., {Tavagnacco}, D., {Terenzi}, L., {Toffolatti}, L., {Tomasi}, M.,
  {Tristram}, M., {Tucci}, M., {T{\"u}rler}, M., {Valenziano}, L., {Valiviita},
  J., {Van Tent}, B., {Vielva}, P., {Villa}, F., {Vittorio}, N., {Wade}, L.~A.,
  {Wandelt}, B.~D., {White}, M., {Yvon}, D., {Zacchei}, A., {Zibin}, J.~P., and
  {Zonca}, A. (2014).
\newblock {Planck 2013 results. XXVII. Doppler boosting of the CMB: Eppur si
  muove}.
\newblock {\em \aap}, 571:A27.

\bibitem[{Price} et~al., 2020]{Price20}
{Price}, D.~C., {Enriquez}, J.~E., {Brzycki}, B., {Croft}, S., {Czech}, D.,
  {DeBoer}, D., {DeMarines}, J., {Foster}, G., {Gajjar}, V., {Gizani}, N.,
  {Hellbourg}, G., {Isaacson}, H., {Lacki}, B., {Lebofsky}, M., {MacMahon}, D.
  H.~E., {Pater}, I.~d., {Siemion}, A. P.~V., {Werthimer}, D., {Green}, J.~A.,
  {Kaczmarek}, J.~F., {Maddalena}, R.~J., {Mader}, S., {Drew}, J., and
  {Worden}, S.~P. (2020).
\newblock {The Breakthrough Listen Search for Intelligent Life: Observations of
  1327 Nearby Stars Over 1.10--3.45 GHz}.
\newblock {\em \aj}, 159(3):86.

\bibitem[{Reid} et~al., 2014]{Reid14}
{Reid}, M.~J., {Menten}, K.~M., {Brunthaler}, A., {Zheng}, X.~W., {Dame},
  T.~M., {Xu}, Y., {Wu}, Y., {Zhang}, B., {Sanna}, A., {Sato}, M., {Hachisuka},
  K., {Choi}, Y.~K., {Immer}, K., {Moscadelli}, L., {Rygl}, K.~L.~J., and
  {Bartkiewicz}, A. (2014).
\newblock {Trigonometric Parallaxes of High Mass Star Forming Regions: The
  Structure and Kinematics of the Milky Way}.
\newblock {\em \apj}, 783(2):130.

\bibitem[Schelling, 1960]{schelling1960strategy}
Schelling, T. (1960).
\newblock {\em The strategy of conflict}.
\newblock Galaxy book. Harvard University Press.

\bibitem[{Sch{\"o}nrich} et~al., 2010]{Schonrich10}
{Sch{\"o}nrich}, R., {Binney}, J., and {Dehnen}, W. (2010).
\newblock {Local kinematics and the local standard of rest}.
\newblock {\em \mnras}, 403(4):1829--1833.

\bibitem[{Sheikh} et~al., 2020]{SheikhETZ}
{Sheikh}, S.~Z., {Siemion}, A., {Enriquez}, J.~E., {Price}, D.~C., {Isaacson},
  H., {Lebofsky}, M., {Gajjar}, V., and {Kalas}, P. (2020).
\newblock {The Breakthrough Listen Search for Intelligent Life: A 3.95-8.00 GHz
  Search for Radio Technosignatures in the Restricted Earth Transit Zone}.
\newblock {\em \aj}, 160(1):29.

\bibitem[{Sheikh} et~al., 2019]{Sheikh19}
{Sheikh}, S.~Z., {Wright}, J.~T., {Siemion}, A., and {Enriquez}, J.~E. (2019).
\newblock {Choosing a Maximum Drift Rate in a SETI Search: Astrophysical
  Considerations}.
\newblock {\em \apj}, 884(1):14.

\bibitem[{Siemion} et~al., 2013]{siemion13}
{Siemion}, A.~P.~V., {Demorest}, P., {Korpela}, E., {Maddalena}, R.~J.,
  {Werthimer}, D., {Cobb}, J., {Howard}, A.~W., {Langston}, G., {Lebofsky}, M.,
  {Marcy}, G.~W., and {Tarter}, J. (2013).
\newblock {A 1.1-1.9 GHz SETI Survey of the Kepler Field. I. A Search for
  Narrow-band Emission from Select Targets}.
\newblock {\em \apj}, 767:94.

\bibitem[{Sivaram} et~al., 2019]{Sivaram2019}
{Sivaram}, C., {Arun}, K., and {Kiren}, O.~V. (2019).
\newblock {Alternative standard frequencies for interstellar communication}.
\newblock {\em International Journal of Astrobiology}, 18(3):209--210.

\bibitem[{Tarter} et~al., 1980]{Tarter1980}
{Tarter}, J.~C., {Cuzzi}, J., {Black}, D., and {Clark}, T. (1980).
\newblock {A high-sensitivity search for extraterrestrial intelligence at
  lambda 18 CM}.
\newblock {\em \icarus}, 42:136--144.

\bibitem[{Tellis} and {Marcy}, 2015]{Tellis15}
{Tellis}, N.~K. and {Marcy}, G.~W. (2015).
\newblock {A Search for Optical Laser Emission Using Keck HIRES}.
\newblock {\em \pasp}, 127(952):540.

\bibitem[{Tellis} and {Marcy}, 2017]{Tellis17}
{Tellis}, N.~K. and {Marcy}, G.~W. (2017).
\newblock {A Search for Laser Emission with Megawatt Thresholds from 5600 FGKM
  Stars}.
\newblock {\em \aj}, 153(6):251.

\bibitem[{Weber}, 1995]{Weber95}
{Weber}, A. (1995).
\newblock {A Biochemical Magic Frequency Based on the Reduction Level of
  Biological Carbon}.
\newblock In {Shostak}, G.~S., editor, {\em Progress in the Search for
  Extraterrestrial Life.}, volume~74 of {\em Astronomical Society of the
  Pacific Conference Series}, page 479.

\bibitem[{Wright}, 2017]{WrightExoplanetsSETI}
{Wright}, J.~T. (2017).
\newblock {Exoplanets and SETI}.
\newblock In {\em Handbook of Exoplanets, Edited by Hans J.~Deeg and Juan
  Antonio Belmonte.~Springer Living Reference Work, ISBN: 978-3-319-30648-3,
  2017, id.186}, page 186. Springer.

\bibitem[{Wright} and {Eastman}, 2014]{bary}
{Wright}, J.~T. and {Eastman}, J.~D. (2014).
\newblock {Barycentric Corrections at 1 cm s$^{-1}$ for Precise Doppler
  Velocities}.
\newblock {\em \pasp}, 126:838--852.

\bibitem[{Wright} et~al., 2018a]{Haystack}
{Wright}, J.~T., {Kanodia}, S., and {Lubar}, E. (2018a).
\newblock {How Much SETI Has Been Done? Finding Needles in the n-dimensional
  Cosmic Haystack}.
\newblock {\em \aj}, 156(6):260.

\bibitem[{Wright} et~al., 2018b]{WrightAdHoc}
{Wright}, J.~T., {Sheikh}, S., {Alm{\'a}r}, I., {Denning}, K., {Dick}, S., and
  {Tarter}, J. (2018b).
\newblock {Recommendations from the Ad Hoc Committee on SETI Nomenclature}.
\newblock {\em arXiv e-prints}, page arXiv:1809.06857.

\end{thebibliography}

\clearpage
\appendix
\setlength{\tabcolsep}{4pt}
\setlength{\extrarowheight}{0pt}


\begin{longtable}{rrrl}
\caption{\label{tab:halpha}The Planck Frequency Comb with $\hbar\rightarrow h$. $n_{h,\alpha} = \log{(E/\sqrt{hc^5/G})}/\log{(e^2/(ch))}$. This set contains no convenient lines in the optical or near-infrared.}\\
\hline
\hline
$n$ & Wavelength & Energy/frequency & Notes\\
\hline
\endfirsthead

\multicolumn{4}{c}%
{{\tablename\ \thetable{} -- continued from previous page}} \\
\hline
\hline
$n$ & Wavelength & Energy/frequency & Notes\\
\hline
\endhead

\hline
\multicolumn{4}{c}%
{{continued on next page}}
\endfoot

\hline
\hline
\endlastfoot

0 &  & $3.06032 \times 10^{28}$ $\mathrm{eV}$  &   \\
\vdots & \vdots & \vdots & \\
13 &  & 121.131 $\mathrm{GeV}$  &  Observable by {\it Fermi} \\
14 &  & 5.55392 $\mathrm{GeV}$  &  Observable by {\it Fermi} \\
15 &  & 254.651 $\mathrm{MeV}$  &  Observable by {\it Fermi} \\
16 &  & 11.6759 $\mathrm{MeV}$  &  Observable by {\it Fermi} \\
17 &  & 535.347 $\mathrm{keV}$  &  Observable by {\it Fermi} \\
18 &  & 24.5460 $\mathrm{keV}$  &  Observable by {\it Fermi} \\
19 & 11.0164 $\mathrm{\mathring{A}}$ & 1.12545 $\mathrm{keV}$  &  Observable by {\it Chandra} \\
20 & 240.268 $\mathrm{\mathring{A}}$ & 51.6025 $\mathrm{eV}$  &   \\
21 & 5240.23 $\mathrm{\mathring{A}}$ & 2.36601 $\mathrm{eV}$  &  Observable by {\it Swift} \\
22 & 11.4289 $\mathrm{\upmu}$ & 26.2310 $\mathrm{THz}$  &  Observable by {\it JWST} \\
23 & 249.264 $\mathrm{\upmu}$ & 1.20271 $\mathrm{THz}$  &   \\
24 & 5.43644 $\mathrm{mm}$ & 55.1450 $\mathrm{GHz}$  &  Microwave U band \\
25 & 11.8569 $\mathrm{cm}$ & 2.52843 $\mathrm{GHz}$  &  Microwave S band \\
26 & 2.58598 $\mathrm{m}$ & 115.930 $\mathrm{MHz}$  &  Observable by EoR experiments \\
27 & 56.4000 $\mathrm{m}$ & 5.31547 $\mathrm{MHz}$  &  Ionospheric cutoff \\
28 &  & 243.717 $\mathrm{kHz}$  &   \\
29 &  & 11.1746 $\mathrm{kHz}$  &   \\
30 &  & 512.362 $\mathrm{Hz}$  &  In the frequency range of LIGO \\
31 &  & 23.4921 $\mathrm{Hz}$  &  In the frequency range of LIGO \\
32 &  & 1.07713 $\mathrm{Hz}$  &   \\
33 &  & 49.3870 $\mathrm{mHz}$  &  In the frequency range of LISA \\
34 &  & 2.26442 $\mathrm{mHz}$  &  In the frequency range of LISA \\
35 &  & 103.825 $\mathrm{\upmu Hz}$  &  In the frequency range of LISA \\
36 &  & 4.76045 $\mathrm{\upmu Hz}$  &   \\
37 &  & 218.270 $\mathrm{nHz}$  &  In the frequency range of PTAs \\
38 &  & 10.0078 $\mathrm{nHz}$  &  In the frequency range of PTAs \\
39 &  & 0.458864 $\mathrm{nHz}$  &   \\

\end{longtable}


\begin{longtable}{rrrl}
\caption{\label{tab:he}The Planck Frequency Comb with $\hbar\rightarrow h$ and base $e$. $n_{h,e} = -\ln{(E/\sqrt{hc^5/G})}$.}\\
\hline
\hline
$n$ & Wavelength & Energy/frequency & Notes\\
\hline
\endfirsthead

\multicolumn{4}{c}%
{{\tablename\ \thetable{} -- continued from previous page}} \\
\hline
\hline
$n$ & Wavelength & Energy/frequency & Notes\\
\hline
\endhead

\hline
\multicolumn{4}{c}%
{{continued on next page}}
\endfoot

\hline
\hline
\endlastfoot

0 &  & $3.06032 \times 10^{28}$ $\mathrm{eV}$  &   \\
\vdots & \vdots & \vdots & \\
39 &  & 353.412 $\mathrm{GeV}$  &   \\
40 &  & 130.013 $\mathrm{GeV}$  &  Observable by {\it Fermi} \\
41 &  & 47.8292 $\mathrm{GeV}$  &  Observable by {\it Fermi} \\
42 &  & 17.5954 $\mathrm{GeV}$  &  Observable by {\it Fermi} \\
43 &  & 6.47297 $\mathrm{GeV}$  &  Observable by {\it Fermi} \\
44 &  & 2.38127 $\mathrm{GeV}$  &  Observable by {\it Fermi} \\
45 &  & 876.022 $\mathrm{MeV}$  &  Observable by {\it Fermi} \\
46 &  & 322.270 $\mathrm{MeV}$  &  Observable by {\it Fermi} \\
47 &  & 118.557 $\mathrm{MeV}$  &  Observable by {\it Fermi} \\
48 &  & 43.6146 $\mathrm{MeV}$  &  Observable by {\it Fermi} \\
49 &  & 16.0449 $\mathrm{MeV}$  &  Observable by {\it Fermi} \\
50 &  & 5.90259 $\mathrm{MeV}$  &  Observable by {\it Fermi} \\
51 &  & 2.17144 $\mathrm{MeV}$  &  Observable by {\it Fermi} \\
52 &  & 798.829 $\mathrm{keV}$  &  Observable by {\it Fermi} \\
53 &  & 293.873 $\mathrm{keV}$  &  Observable by {\it Fermi} \\
54 &  & 108.110 $\mathrm{keV}$  &  Observable by {\it Fermi} \\
55 &  & 39.7713 $\mathrm{keV}$  &  Observable by {\it Fermi} \\
56 &  & 14.6311 $\mathrm{keV}$  &  Observable by {\it Fermi} \\
57 & 2.30348 $\mathrm{\mathring{A}}$ & 5.38246 $\mathrm{keV}$  &  Observable by {\it Chandra} \\
58 & 6.26152 $\mathrm{\mathring{A}}$ & 1.98010 $\mathrm{keV}$  &  Observable by {\it Chandra} \\
59 & 17.0206 $\mathrm{\mathring{A}}$ & 728.437 $\mathrm{eV}$  &  Observable by {\it Chandra} \\
60 & 46.2667 $\mathrm{\mathring{A}}$ & 267.977 $\mathrm{eV}$  &  Observable by {\it Chandra} \\
61 & 125.766 $\mathrm{\mathring{A}}$ & 98.5833 $\mathrm{eV}$  &   \\
62 & 341.867 $\mathrm{\mathring{A}}$ & 36.2668 $\mathrm{eV}$  &   \\
63 & 929.292 $\mathrm{\mathring{A}}$ & 13.3418 $\mathrm{eV}$  &   \\
64 & 2526.08 $\mathrm{\mathring{A}}$ & 4.90817 $\mathrm{eV}$  &   \\
65 & 6866.59 $\mathrm{\mathring{A}}$ & 1.80562 $\mathrm{eV}$  &  Optical \\
66 & 1.86653 $\mathrm{\upmu}$ & 160.615 $\mathrm{THz}$  &  Infrared H band \\
67 & 5.07376 $\mathrm{\upmu}$ & 59.0868 $\mathrm{THz}$  &  Observable by {\it JWST} \\
68 & 13.7919 $\mathrm{\upmu}$ & 21.7368 $\mathrm{THz}$  &  Observable by {\it JWST} \\
69 & 37.4903 $\mathrm{\upmu}$ & 7.99653 $\mathrm{THz}$  &  Infrared Z band \\
70 & 101.909 $\mathrm{\upmu}$ & 2.94176 $\mathrm{THz}$  &   \\
71 & 277.018 $\mathrm{\upmu}$ & 1.08221 $\mathrm{THz}$  &   \\
72 & 753.013 $\mathrm{\upmu}$ & 398.124 $\mathrm{GHz}$  &  Observable with ALMA \\
73 & 2.04690 $\mathrm{mm}$ & 146.462 $\mathrm{GHz}$  &  Observable with ALMA \\
74 & 5.56405 $\mathrm{mm}$ & 53.8802 $\mathrm{GHz}$  &  Microwave U band \\
75 & 1.51247 $\mathrm{cm}$ & 19.8214 $\mathrm{GHz}$  &  Microwave K band \\
76 & 4.11131 $\mathrm{cm}$ & 7.29189 $\mathrm{GHz}$  &  Microwave C band \\
77 & 11.1757 $\mathrm{cm}$ & 2.68254 $\mathrm{GHz}$  &  Microwave S band \\
78 & 30.3787 $\mathrm{cm}$ & 986.851 $\mathrm{MHz}$  &   \\
79 & 82.5779 $\mathrm{cm}$ & 363.042 $\mathrm{MHz}$  &   \\
80 & 2.24470 $\mathrm{m}$ & 133.556 $\mathrm{MHz}$  &  Observable by EoR experiments \\
81 & 6.10173 $\mathrm{m}$ & 49.1324 $\mathrm{MHz}$  &  Observable by EoR experiments \\
82 & 16.5862 $\mathrm{m}$ & 18.0748 $\mathrm{MHz}$  &   \\
83 & 45.0860 $\mathrm{m}$ & 6.64935 $\mathrm{MHz}$  &  Ionospheric cutoff \\
84 & 122.556 $\mathrm{m}$ & 2.44616 $\mathrm{MHz}$  &   \\
85 & 333.143 $\mathrm{m}$ & 899.891 $\mathrm{kHz}$  &   \\
86 & 905.576 $\mathrm{m}$ & 331.052 $\mathrm{kHz}$  &   \\
87 &  & 121.787 $\mathrm{kHz}$  &   \\
88 &  & 44.8029 $\mathrm{kHz}$  &   \\
89 &  & 16.4821 $\mathrm{kHz}$  &   \\
90 &  & 6.06342 $\mathrm{kHz}$  &  In the frequency range of LIGO \\
91 &  & 2.23061 $\mathrm{kHz}$  &  In the frequency range of LIGO \\
92 &  & 820.595 $\mathrm{Hz}$  &  In the frequency range of LIGO \\
93 &  & 301.880 $\mathrm{Hz}$  &  In the frequency range of LIGO \\
94 &  & 111.055 $\mathrm{Hz}$  &  In the frequency range of LIGO \\
95 &  & 40.8550 $\mathrm{Hz}$  &  In the frequency range of LIGO \\
96 &  & 15.0297 $\mathrm{Hz}$  &  In the frequency range of LIGO \\
97 &  & 5.52912 $\mathrm{Hz}$  &   \\
98 &  & 2.03405 $\mathrm{Hz}$  &   \\
99 &  & 748.285 $\mathrm{mHz}$  &   \\
100 &  & 275.279 $\mathrm{mHz}$  &   \\
101 &  & 101.269 $\mathrm{mHz}$  &   \\
102 &  & 37.2549 $\mathrm{mHz}$  &  In the frequency range of LISA \\
103 &  & 13.7053 $\mathrm{mHz}$  &  In the frequency range of LISA \\
104 &  & 5.04191 $\mathrm{mHz}$  &  In the frequency range of LISA \\
105 &  & 1.85481 $\mathrm{mHz}$  &  In the frequency range of LISA \\
106 &  & 682.348 $\mathrm{\upmu Hz}$  &  In the frequency range of LISA \\
107 &  & 251.022 $\mathrm{\upmu Hz}$  &  In the frequency range of LISA \\
108 &  & 92.3458 $\mathrm{\upmu Hz}$  &   \\
109 &  & 33.9721 $\mathrm{\upmu Hz}$  &   \\
110 &  & 12.4976 $\mathrm{\upmu Hz}$  &   \\
111 &  & 4.59762 $\mathrm{\upmu Hz}$  &   \\
112 &  & 1.69137 $\mathrm{\upmu Hz}$  &   \\
113 &  & 622.221 $\mathrm{nHz}$  &  In the frequency range of PTAs \\
114 &  & 228.902 $\mathrm{nHz}$  &  In the frequency range of PTAs \\
115 &  & 84.2084 $\mathrm{nHz}$  &  In the frequency range of PTAs \\
116 &  & 30.9786 $\mathrm{nHz}$  &  In the frequency range of PTAs \\
117 &  & 11.3964 $\mathrm{nHz}$  &  In the frequency range of PTAs \\
118 &  & 4.19249 $\mathrm{nHz}$  &  In the frequency range of PTAs \\
119 &  & 1.54233 $\mathrm{nHz}$  &  In the frequency range of PTAs \\
120 &  & 0.567392 $\mathrm{nHz}$  &   \\

\end{longtable}


\begin{longtable}{rrrl}
\caption{\label{tab:hbare}The Planck Frequency Comb with base $e$. $n_{\hbar,e} = -\ln{(E/E_P)}$.}\\
\hline
\hline
$n$ & Wavelength & Energy/frequency & Notes\\
\hline
\endfirsthead

\multicolumn{4}{c}%
{{\tablename\ \thetable{} -- continued from previous page}} \\
\hline
\hline
$n$ & Wavelength & Energy/frequency & Notes\\
\hline
\endhead

\hline
\multicolumn{4}{c}%
{{continued on next page}}
\endfoot

\hline
\hline
\endlastfoot

0 &  & $1.22089 \times 10^{28}$ $\mathrm{eV}$  &  Planck energy \\
\vdots & \vdots & \vdots & \\
39 &  & 140.991 $\mathrm{GeV}$  &  Observable by {\it Fermi} \\
40 &  & 51.8678 $\mathrm{GeV}$  &  Observable by {\it Fermi} \\
41 &  & 19.0811 $\mathrm{GeV}$  &  Observable by {\it Fermi} \\
42 &  & 7.01954 $\mathrm{GeV}$  &  Observable by {\it Fermi} \\
43 &  & 2.58234 $\mathrm{GeV}$  &  Observable by {\it Fermi} \\
44 &  & 949.991 $\mathrm{MeV}$  &  Observable by {\it Fermi} \\
45 &  & 349.482 $\mathrm{MeV}$  &  Observable by {\it Fermi} \\
46 &  & 128.567 $\mathrm{MeV}$  &  Observable by {\it Fermi} \\
47 &  & 47.2973 $\mathrm{MeV}$  &  Observable by {\it Fermi} \\
48 &  & 17.3997 $\mathrm{MeV}$  &  Observable by {\it Fermi} \\
49 &  & 6.40099 $\mathrm{MeV}$  &  Observable by {\it Fermi} \\
50 &  & 2.35479 $\mathrm{MeV}$  &  Observable by {\it Fermi} \\
51 &  & 866.280 $\mathrm{keV}$  &  Observable by {\it Fermi} \\
52 &  & 318.686 $\mathrm{keV}$  &  Observable by {\it Fermi} \\
53 &  & 117.238 $\mathrm{keV}$  &  Observable by {\it Fermi} \\
54 &  & 43.1295 $\mathrm{keV}$  &  Observable by {\it Fermi} \\
55 &  & 15.8665 $\mathrm{keV}$  &  Observable by {\it Fermi} \\
56 & 2.12413 $\mathrm{\mathring{A}}$ & 5.83695 $\mathrm{keV}$  &  Observable by {\it Chandra} \\
57 & 5.77398 $\mathrm{\mathring{A}}$ & 2.14729 $\mathrm{keV}$  &  Observable by {\it Chandra} \\
58 & 15.6953 $\mathrm{\mathring{A}}$ & 789.945 $\mathrm{eV}$  &  Observable by {\it Chandra} \\
59 & 42.6642 $\mathrm{\mathring{A}}$ & 290.604 $\mathrm{eV}$  &  Observable by {\it Chandra} \\
60 & 115.973 $\mathrm{\mathring{A}}$ & 106.907 $\mathrm{eV}$  &  Observable by {\it Chandra} \\
61 & 315.248 $\mathrm{\mathring{A}}$ & 39.3290 $\mathrm{eV}$  &   \\
62 & 856.934 $\mathrm{\mathring{A}}$ & 14.4683 $\mathrm{eV}$  &   \\
63 & 2329.39 $\mathrm{\mathring{A}}$ & 5.32261 $\mathrm{eV}$  &   \\
64 & 6331.94 $\mathrm{\mathring{A}}$ & 1.95808 $\mathrm{eV}$  &  Optical \\
65 & 1.72120 $\mathrm{\upmu}$ & 174.177 $\mathrm{THz}$  &  Infrared H band \\
66 & 4.67870 $\mathrm{\upmu}$ & 64.0760 $\mathrm{THz}$  &  Observable by {\it JWST} \\
67 & 12.7180 $\mathrm{\upmu}$ & 23.5722 $\mathrm{THz}$  &  Observable by {\it JWST} \\
68 & 34.5712 $\mathrm{\upmu}$ & 8.67174 $\mathrm{THz}$  &  Infrared Z band \\
69 & 93.9743 $\mathrm{\upmu}$ & 3.19016 $\mathrm{THz}$  &   \\
70 & 255.449 $\mathrm{\upmu}$ & 1.17359 $\mathrm{THz}$  &   \\
71 & 694.381 $\mathrm{\upmu}$ & 431.741 $\mathrm{GHz}$  &  Observable with ALMA \\
72 & 1.88752 $\mathrm{mm}$ & 158.828 $\mathrm{GHz}$  &  Observable with ALMA \\
73 & 5.13082 $\mathrm{mm}$ & 58.4297 $\mathrm{GHz}$  &  Microwave U band \\
74 & 1.39470 $\mathrm{cm}$ & 21.4951 $\mathrm{GHz}$  &  Microwave K band \\
75 & 3.79119 $\mathrm{cm}$ & 7.90760 $\mathrm{GHz}$  &  Microwave C band \\
76 & 10.3055 $\mathrm{cm}$ & 2.90905 $\mathrm{GHz}$  &  Microwave S band \\
77 & 28.0133 $\mathrm{cm}$ & 1.07018 $\mathrm{GHz}$  &  Microwave L band \\
78 & 76.1481 $\mathrm{cm}$ & 393.696 $\mathrm{MHz}$  &   \\
79 & 2.06992 $\mathrm{m}$ & 144.833 $\mathrm{MHz}$  &  Observable by EoR experiments \\
80 & 5.62663 $\mathrm{m}$ & 53.2810 $\mathrm{MHz}$  &  Observable by EoR experiments \\
81 & 15.2948 $\mathrm{m}$ & 19.6010 $\mathrm{MHz}$  &   \\
82 & 41.5755 $\mathrm{m}$ & 7.21080 $\mathrm{MHz}$  &  Ionospheric cutoff \\
83 & 113.014 $\mathrm{m}$ & 2.65271 $\mathrm{MHz}$  &   \\
84 & 307.203 $\mathrm{m}$ & 975.876 $\mathrm{kHz}$  &   \\
85 & 835.066 $\mathrm{m}$ & 359.005 $\mathrm{kHz}$  &   \\
86 &  & 132.070 $\mathrm{kHz}$  &   \\
87 &  & 48.5860 $\mathrm{kHz}$  &   \\
88 &  & 17.8738 $\mathrm{kHz}$  &   \\
89 &  & 6.57540 $\mathrm{kHz}$  &  In the frequency range of LIGO \\
90 &  & 2.41895 $\mathrm{kHz}$  &  In the frequency range of LIGO \\
91 &  & 889.884 $\mathrm{Hz}$  &  In the frequency range of LIGO \\
92 &  & 327.370 $\mathrm{Hz}$  &  In the frequency range of LIGO \\
93 &  & 120.433 $\mathrm{Hz}$  &  In the frequency range of LIGO \\
94 &  & 44.3047 $\mathrm{Hz}$  &  In the frequency range of LIGO \\
95 &  & 16.2988 $\mathrm{Hz}$  &  In the frequency range of LIGO \\
96 &  & 5.99599 $\mathrm{Hz}$  &   \\
97 &  & 2.20580 $\mathrm{Hz}$  &   \\
98 &  & 811.469 $\mathrm{mHz}$  &   \\
99 &  & 298.523 $\mathrm{mHz}$  &   \\
100 &  & 109.820 $\mathrm{mHz}$  &   \\
101 &  & 40.4007 $\mathrm{mHz}$  &  In the frequency range of LISA \\
102 &  & 14.8626 $\mathrm{mHz}$  &  In the frequency range of LISA \\
103 &  & 5.46763 $\mathrm{mHz}$  &  In the frequency range of LISA \\
104 &  & 2.01143 $\mathrm{mHz}$  &  In the frequency range of LISA \\
105 &  & 739.964 $\mathrm{\upmu Hz}$  &  In the frequency range of LISA \\
106 &  & 272.217 $\mathrm{\upmu Hz}$  &  In the frequency range of LISA \\
107 &  & 100.143 $\mathrm{\upmu Hz}$  &  In the frequency range of LISA \\
108 &  & 36.8406 $\mathrm{\upmu Hz}$  &   \\
109 &  & 13.5529 $\mathrm{\upmu Hz}$  &   \\
110 &  & 4.98584 $\mathrm{\upmu Hz}$  &   \\
111 &  & 1.83419 $\mathrm{\upmu Hz}$  &   \\
112 &  & 674.760 $\mathrm{nHz}$  &  In the frequency range of PTAs \\
113 &  & 248.230 $\mathrm{nHz}$  &  In the frequency range of PTAs \\
114 &  & 91.3188 $\mathrm{nHz}$  &  In the frequency range of PTAs \\
115 &  & 33.5943 $\mathrm{nHz}$  &  In the frequency range of PTAs \\
116 &  & 12.3587 $\mathrm{nHz}$  &  In the frequency range of PTAs \\
117 &  & 4.54649 $\mathrm{nHz}$  &  In the frequency range of PTAs \\
118 &  & 1.67256 $\mathrm{nHz}$  &  In the frequency range of PTAs \\
119 &  & 0.615301 $\mathrm{nHz}$  &   \\

\end{longtable}

\clearpage
\begin{longtable}{rrrl}
\caption{\label{tab:Ry}The Rydberg Frequency Comb. $n_R = \log_\alpha{(E/(m_ec^2/2))}$}\\
\hline
\hline
$n$ & Wavelength & Energy/frequency & Notes\\
\hline
\endfirsthead

\multicolumn{4}{c}%
{{\tablename\ \thetable{} -- continued from previous page}} \\
\hline
\hline
$n$ & Wavelength & Energy/frequency & Notes\\
\hline
\endhead

\hline
\multicolumn{4}{c}%
{{continued on next page}}
\endfoot

\hline
\hline
\endlastfoot

-2 &  & 4.79799 $\mathrm{GeV}$  &  Observable by {\it Fermi} \\
-1 &  & 35.0126 $\mathrm{MeV}$  &  Observable by {\it Fermi} \\
0 &  & 255.499 $\mathrm{keV}$  &  Observable by {\it Fermi} \\
1 & 6.64984 $\mathrm{\mathring{A}}$ & 1.86447 $\mathrm{keV}$  &  Observable by {\it Chandra} \\
2 & 911.267 $\mathrm{\mathring{A}}$ & 13.6057 $\mathrm{eV}$  &  Rydberg \\
3 & 12.4876 $\mathrm{\upmu}$ & 24.0071 $\mathrm{THz}$  &  Observable by {\it JWST} \\
4 & 1.71126 $\mathrm{mm}$ & 175.189 $\mathrm{GHz}$  &  Observable with ALMA \\
5 & 23.4504 $\mathrm{cm}$ & 1.27841 $\mathrm{GHz}$  &  Microwave L band \\
6 & 32.1354 $\mathrm{m}$ & 9.32903 $\mathrm{MHz}$  &   \\
7 &  & 68.0772 $\mathrm{kHz}$  &   \\
8 &  & 496.783 $\mathrm{Hz}$  &  In the frequency range of LIGO \\
9 &  & 3.62520 $\mathrm{Hz}$  &   \\
10 &  & 26.4544 $\mathrm{mHz}$  &  In the frequency range of LISA \\
11 &  & 193.047 $\mathrm{\upmu Hz}$  &  In the frequency range of LISA \\
12 &  & 1.40873 $\mathrm{\upmu Hz}$  &   \\
13 &  & 10.2800 $\mathrm{nHz}$  &  In the frequency range of PTAs \\

\end{longtable}

\end{document}